\documentclass[useAMS,usenatbib]{mn2e}

\usepackage{graphicx}
\usepackage{natbib}

\catcode`\@=11
\def\gsim{\ifmmode{\mathrel{\mathpalette\@versim>}}
    \else{$\mathrel{\mathpalette\@versim>}$}\fi}
\def\lsim{\ifmmode{\mathrel{\mathpalette\@versim<}}
    \else{$\mathrel{\mathpalette\@versim<}$}\fi}
\def\@versim#1#2{\lower 2.9truept \vbox{\baselineskip 0pt \lineskip
    0.5truept \ialign{$\m@th#1\hfil##\hfil$\crcr#2\crcr\sim\crcr}}}
\catcode`\@=12

\newcommand{\beq}{\begin{equation}}
\newcommand{\eeq}{\end{equation}}
\newcommand{\azero}{{a_0}}
\newcommand{\rtilde}{\tilde{r}}
\def\NMODY{{\sc n-mody}}

\newcommand{\gv}{{\bf g}}

\newcommand{\xv}{{\bf x}}
\newcommand{\ve}{v_{\rm e}}
\newcommand{\rt}{r_{\rm t}}
\newcommand{\fzero}{f_0}
\newcommand{\rc}{r_{\rm c}}
\newcommand{\ra}{r_{\rm a}}
\newcommand{\LV}{L_V}
\newcommand{\phiext}{\phi_{\rm ext}}
\newcommand{\gext}{g_{\rm ext}}
\newcommand{\gvext}{{\bf g}_{\rm ext}}
\newcommand{\sigmaK}{\sigma_{\rm K}}
\newcommand{\sigmaKsq}{\sigma_{\rm K}^2}
\newcommand{\sigmav}{\sigma_{\rm v}}
\newcommand{\sigmavsq}{\sigma_{\rm v}^2}
\newcommand{\sigmaLOS}{\sigma_{\rm LOS}}
\newcommand{\sigmaLOSsq}{\sigma_{\rm LOS}^2}
\def\kms{{\rm \,km\,s^{-1}}}
\def\ms{{\rm \,m\,s^{-1}}}
\def\d{{\rm d}}

\title[Globular clusters in MOND]{Globular clusters in modified Newtonian dynamics: velocity-dispersion profiles from self-consistent models} 

\author[A. Sollima and C. Nipoti]{A. Sollima$^{1}$\thanks{E-mail:
    asollima@iac.es (AS)} and C.  Nipoti$^{2}$\\ $^{1}$Instituto de
  Astrofisica de Canarias, C/Via Lactea s/n, E-38205, San Cristobal de
  La Laguna, Tenerife, Spain\\ $^{2}$Dipartimento di Astronomia, Universit\`a di Bologna, Via
  Ranzani 1, I-40127, Bologna, Italy}
\begin{document}

\date{Revised draft, August 24, 2009}

\pagerange{\pageref{firstpage}--\pageref{lastpage}} \pubyear{2009}

\maketitle

\label{firstpage}

\begin{abstract}
  We test the modified Newtonian
  dynamics (MOND) theory with the velocity-dispersion profiles of
  Galactic globular clusters populating the outermost region of the
  Milky Way halo, where the Galactic acceleration is lower than the
  characteristic MOND acceleration $\azero$.  For this purpose, we
  constructed self-consistent, spherical models of stellar systems in
  MOND, which are the analogues of the Newtonian King models. The
  models are spatially limited, reproduce well the surface-brightness
  profiles of globular clusters, and have velocity-dispersion profiles
  that differ remarkably in shape from the corresponding Newtonian
  models. We present dynamical models of six globular clusters,
  which can be used to efficiently test MOND with the available
  observing facilities.  A comparison with recent spectroscopic data
  obtained for NGC2419 suggests that the kinematics of this cluster
  might be hard to explain in MOND.
\end{abstract}


\section{Introduction}
\label{intro}

Modified Newtonian dynamics (MOND; Milgrom 1983) represents one of the
most popular alternative to the commonly accepted dark-matter
paradigm.  This theory was put forward in the early eighties as a way to
explain the observed flat rotation curves of disk galaxies without the
need of hidden matter. In Bekenstein \& Milgrom's (1984) formulation
of MOND, Poisson's equation $\nabla^2\phi_{\rm N}=4\pi G\rho$ is
replaced by the field equation
 \begin{equation}
\nabla\cdot\left[\mu\left({\Vert\nabla\phi\Vert\over\azero}\right)
\nabla\phi\right] = 4\pi G \rho,
\label{eqMOND}
\end{equation} 
where $\Vert ...\Vert$ is the standard Euclidean norm, and $\phi$ is
the gravitational potential for MOND. The gravitational acceleration
is $\gv=-\nabla\phi$ just as the Newtonian acceleration
$\gv_{\rm N}=-\nabla\phi_{\rm N}$.  For a system of finite mass,
$\Vert\nabla\phi\Vert\to 0$ as $\Vert\xv\Vert\to\infty$, where $\xv$
is the position vector relative to the system's centre of mass.  The
function $\mu(y)$ is constrained by the theory only to the extent that
it must run smoothly from $\mu(y)\sim y$ at $y\ll 1$ (the so-called
``deep-MOND'' regime) to $\mu(y)\sim 1$ at $y\gg 1$ (the Newtonian
regime), with the transition taking place at $y\approx 1$ (i.e., when
$\Vert\nabla\phi\Vert$ is of order of the characteristic acceleration
$\azero \simeq 1.2 \times 10^{-10} {\rm m}\,{\rm s}^{-2}$).  In
spherical symmetry equation~(\ref{eqMOND}) reduces to Milgrom's (1983)
original phenomenological relation
\begin{equation}
\mu\left(\frac{\Vert\gv\Vert}{a_{0}}\right)\gv=\gv_{\rm N},
\label{eqMONDsph}
\end{equation}  
in which $\gv$ and $\gv_{\rm N}$ are parallel, $g \sim g_{\rm N}$ when
$g/\azero\gg 1$ and $g\sim\sqrt{\azero g_{\rm N}}$ when $g/\azero\ll 1$.

Over more than two decades the theory has been quite successful, 
resisting several attempts at falsification (see, e.g. Sanders \& McGaugh 2002;
Bekenstein 2006; Milgrom 2008; Bekenstein 2009).  However, there are
cases in which MOND appears to have difficulties in explaining the
observed data: the X-ray and gravitational lensing
properties of clusters of galaxies (e.g., The \& White 1988; Sanders
2007; Natarajan \& Zhao 2008), and in particular the 
"Bullet" cluster (Clowe et al. 2006; Angus et al. 2007), the internal
dynamics of dwarf spheroidal galaxies (Kleyna et al.~2001;
S\'anchez-Salcedo, Reyes-Iturbide \& Hernandez 2006; Nipoti et
al. 2008; Angus 2008; Angus \& Diaferio 2009), the phenomenon of
galaxy merging (Nipoti, Londrillo \& Ciotti~2007a), and the vertical
kinematics of the Milky Way (Nipoti et al. 2007b; Bienaym{\'e} et
al. 2009). Here we present a further test of MOND using the 
Milky Way Globular Clusters (GCs).

Typical globular clusters are not ideal systems to test the MOND
hypothesis because they are characterized by high stellar-mass
surface density and internal acceleration larger than $\azero$.
However, there are several cases of less dense GCs, with internal
acceleration comparable to or smaller than $\azero$, which represent
good candidates to test MOND (Baumgardt, Grebel \& Kroupa 2005,
hereafter BGK05).
A complication that arise in testing MOND with GCs is the
so-called ``external field effect'' (Bekenstein \& Milgrom~1984).
In MOND the {\it internal} dynamics of a system is
affected by the presence of even an {\it uniform} external
gravitational field $\gvext$.  As a consequence, the interpretation in
the context of MOND of GC dynamics is typically difficult because one
needs to account for the presence of the external gravitational field
due to the Galaxy. An exception is represented by the GCs populating
the outermost region of the Galactic halo, which are far enough to
experience only a negligible acceleration from the Milky Way
($\gext\ll\azero$), and therefore constitute an ideal laboratory to
test MOND.

The application of MOND to GCs has been studied by different
authors by estimating the overall velocity dispersions of GCs through
the virial theorem (e.g. BGK05) or by deriving the their MOND
velocity-dispersion profiles through the Jeans equations (e.g. Moffat
\& Toth 2008).  As in Newtonian gravity, also in MOND the latter
approach has the limitation that it is not based on self-consistent
models, so that it is not guaranteed that there is a non-negative
distribution function corresponding to the assumed density
distribution.  Haghi et al. (2009) solved this problem by obtaining
equilibrium MOND GC models as end-products of N-body simulations run
with the numerical code \NMODY\ (Londrillo \& Nipoti 2008), based on
the MOND potential solver described in Ciotti, Londrillo \& Nipoti
(2006). N-body models are useful when the external field is important
and the GC is not spherically symmetric. When the external field
  is weak, the GC can be well represented by a spherical stellar
  system and it is possible to construct self-consistent models
  without resorting to simulations.

In this paper we present self-consistent, spherically symmetric models
of stellar systems that are MOND steady-state solutions of the
Fokker-Planck equation, which are the analogues of Newtonian King
(1966) models.  We use them to predict the MOND velocity-dispersion
profiles of GCs belonging to the outer Galactic halo, which can be
used to discriminate between MOND and Newtonian gravity. In Section 2
we describe the theoretical basis of our models and discuss their
structural and kinematic properties.  Section 3 is devoted to the
application of the models to a sample of six GCs located in the
external parts of the Galaxy. Finally, we summarize and discuss our
results in Section 4.

\section{Self-consistent spherical models of globular clusters in MOND}
\label{mod_sec}

To describe the phase-space distribution of a spherically
  symmetric star cluster in MOND we choose the steady-state solution
  of the Fokker-Planck equation proposed by King~(1966), but with
  equation~(\ref{eqMOND}) replacing Poisson's equation.  The
  distribution function is
\begin{equation}
f(r,v)=\fzero
\exp\left({-\frac{\psi-\psi_0}{\sigmaK^2}}\right)
\left[
{\exp\left({-\frac{v^{2}}{2 \sigmaK^2}}\right)-\exp\left({-\frac{\ve^{2}}{2 \sigmaK^{2}}}\right)}
\right],
\label{df_eq}
\end{equation}
where the effective potential $\psi$ is the difference between the
cluster potential at a given radius $r$ and the potential outside the
cluster tidal extent $\psi\equiv\phi-\phiext$, $\psi_0$ is the central
effective potential, $\fzero$ is a scale factor, $\ve=\sqrt{-2 \psi}$ is
the cluster escape velocity, and $\sigmaK$ is a normalization term
which is proportional to the central velocity dispersion.  The density
$\rho$ and the 3D velocity dispersion $\sigmav$ can be obtained by
integrating the distribution function:
\begin{eqnarray}
\rho(r)&=& \int_{0}^{\ve} 4 \pi v^{2} ~f(r,v) ~\d v, \\
\sigmavsq(r)&=&\frac{1}{\rho(r)}\int_{0}^{\ve} 4 \pi v^{4}~ f(r,v)
~\d v.
\end{eqnarray}
The above equations can be written in terms of dimensionless quantities
by substituting
\begin{eqnarray}
W=-{\psi \over \sigmaKsq}, & \eta={v^2\over 2
  \sigmaKsq},\\ \tilde{\rho}={\rho\over \rho_{0}}, & \rtilde={r\over\rc},
\end{eqnarray}
where $\rho_{0}=\rho(0)$ is the central cluster density and
\begin{equation}
\label{rc_eq} 
\rc\equiv\left({9 \sigmaKsq \over 4 \pi G \rho_{0}}\right)^{1/2}
\end{equation}
is the core radius (King 1966), so we get
 \begin{equation}
\label{den_eq}
\tilde{\rho}=e^{(W-W_{0})} \frac{\int_{0}^{W} \eta^{\frac{3}{2}}e^{-\eta}~\d\eta}{\int_{0}^{W_{0}} \eta^{\frac{3}{2}}e^{-\eta}~\d\eta}
\end{equation}
and
\begin{equation}
\label{sig_eq}
\sigmavsq=\frac{6}{5} \sigmaKsq\frac{\int_{0}^{W} \eta^{\frac{5}{2}}e^{-\eta}~\d\eta}{\int_{0}^{W}
\eta^{\frac{3}{2}}e^{-\eta}~\d\eta}.
\end{equation}
The above expressions give the density and velocity-dispersion radial
profiles as functions of the dimensionless potential $W(\rtilde)$,
which can be calculated by solving the modified Poisson
equation~(\ref{eqMOND}) with the boundary conditions at the centre
\begin{eqnarray}
\label{bound_eq}
&&W= W_{0}, \nonumber\\
&&\frac{\d W}{\d\rtilde}=0.
\end{eqnarray}
In spherical symmetry equation~(\ref{eqMOND}) can be written in the
form
\begin{equation}
 \frac{1}{r^{2}}\frac{\d}{\d r}\left[r^{2}\mu\left({{1\over\azero}\left|{\d \psi\over \d r}\right|}\right)\frac{\d\psi}{\d r}\right]=4\pi G
\rho
\end{equation}
or, using dimensionless quantities,
\begin{equation}
\label{poiss_eq}
\frac{1}{\rtilde^{2}}\frac{\d}{\d\rtilde}\left[\rtilde^{2}
\mu\left({\xi\left|{\d W \over \d \rtilde}\right|}\right)
\frac{\d W}{\d\rtilde}\right]= -9\tilde{\rho},
\end{equation}
where $\xi\equiv\sigmaKsq / {a_{0}\rc}$ is a dimensionless parameter, which is smaller for systems closer to
  the deep-MOND regime.

For a given choice of the ($W_{0},\xi$) pair, $W(\rtilde)$ has
  been obtained by numerically integrating equation~(\ref{poiss_eq})
  from the centre imposing the boundary conditions indicated in
  equation~(\ref{bound_eq}) (see Appendix).  The corresponding radial
  density and 3D velocity-dispersion profiles are then
  projected on the plane of the sky, giving the line-of-sight (LOS) 
  velocity dispersion $\sigma_\mathrm{LOS}$ as
\begin{equation}
\label{los_eq}
\sigmaLOSsq(R)={2\over\Sigma(R)}\int_R^\infty {\rho(r)~\sigmavsq(r) r \d r\over 3 \sqrt{r^2-R^2}},
\end{equation}
where
\begin{equation}
\Sigma(R)=2\int_R^\infty {\rho(r)r \d r\over\sqrt{r^2-R^2}}
\end{equation}
is the surface density.  The described procedure is straightforward
and produces self-consistent equilibrium models, truncated at a tidal
radius $\rt$.

A particular case of this family of models was studied by Brada
  \& Milgrom~(2000), who constructed steady state King models of
stellar systems in deep-MOND [$\mu(y)=y$] regime as initial conditions
for their N-body simulations of dwarf galaxies. Here we extend their
models to general MOND cases in which the internal acceleration is not
necessarily everywhere small as compared to $\azero$.

\subsection{Limits of validity of the models}
\label{ext_sec}

The models presented in the above Section were constructed neglecting
the effects of an external field on the modified Poisson equation
(\ref{eqMOND}), so some considerations on the limits of validity of
the models are necessary. The MOND field of a system with density
distribution $\rho(r)$ in the presence of the external field $\gvext$
can be obtained by solving equation (\ref{eqMOND}) with boundary
conditions $\nabla\phi\to \gvext$ for $\Vert\xv\Vert\to\infty$
(Bekenstein \& Milgrom 1984). Computing such a field is in general a
difficult task, because the presence of an external field breaks
  the symmetry of the system. However, the departure from spherical
symmetry can be safely neglected in the external acceleration regime
$\gext \ll \azero$ (Milgrom \& Bekenstein 1987), so our spherically
symmetric models are not valid when the external field is comparable
to or larger than $\azero$.
 
Another important issue is the concept of escape velocity in MOND
  models.  In the formulation presented in Section \ref{mod_sec}, the
velocity distribution at a given radius is truncated at the local
escape velocity $\ve=\sqrt{-2\psi}$ (see also Sanchez-Salcedo \&
Hernandez 2007; Wu et al. 2008).  This condition is no longer valid in
the case of an isolated MOND model ($\gext=0$), in which a particle
with arbitrarily large speed remains bound to the cluster.  In a more
realistic case, when the system is surrounded by other objects
($\gext>0$), the total gravitational potential will admit a maximum
value $\phiext$ at a finite distance from the cluster centre (see Wu
et al. 2007). A cluster star located at a distance $r$ from the
  cluster centre, with a speed $v>\ve$, will reach the radius $\rt$
  with non-null velocity, feel the external potential as dominant and
  never return.

Summarizing, on the basis of the above considerations on the external
field effect and on the escape velocity in MOND, our models are
  acceptable approximations of GCs when the external acceleration
field lies in the range $0<\gext\ll\azero$.

\begin{figure}
 \includegraphics[width=8.7cm]{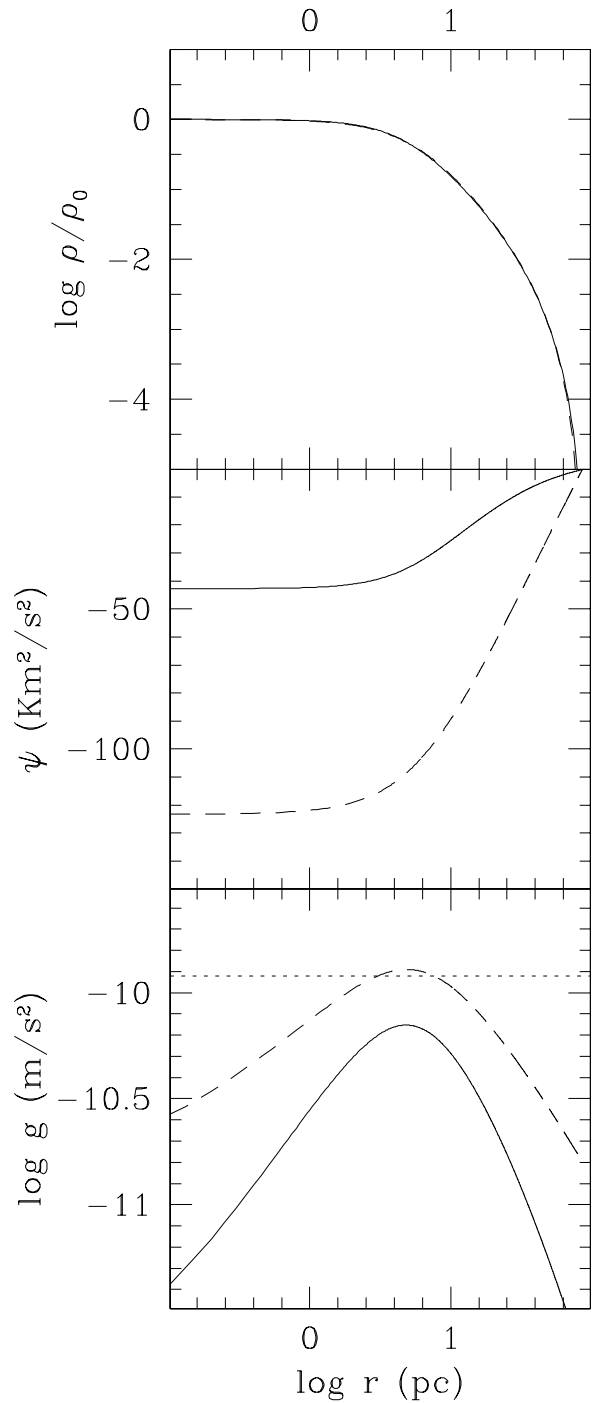}
\caption{Density (top panel), gravitational potential (middle panel)
  and internal acceleration modulus $g$ (bottom panel) as functions of
  the distance from the cluster centre for a Newtonian King model with
  $W_0=6$, $\log(M/M_{\odot})=5$ and $\rc=5$ pc (solid lines) and of a
  MOND model with $W_0=9.5$, $\log(M/M_{\odot})=5$ and $\xi=0.53$
  (dashed lines).}
\label{pot}
\end{figure}

\subsection{Structural and kinematic properties of the models} 
\label{res_sect}

\begin{figure*}
 \includegraphics[width=15.cm]{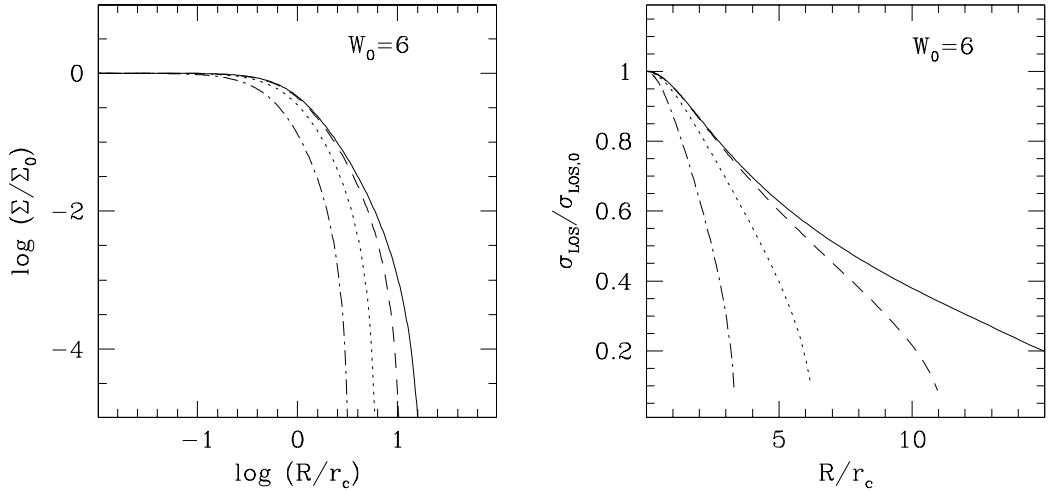}
  \caption{Projected surface-density and LOS velocity-dispersion
    profiles of $W_{0}=6$ MOND models with $\xi=0.1$ (dot-dashed
    lines), $\xi=1$ (dotted lines) and $\xi=10$ (dashed lines).  For
    comparison, the profiles of $W_{0}=6$ King Newtonian models are
    shown in both panels as solid lines.}
\label{xi}
\end{figure*}

\begin{figure*}
 \includegraphics[width=15.cm]{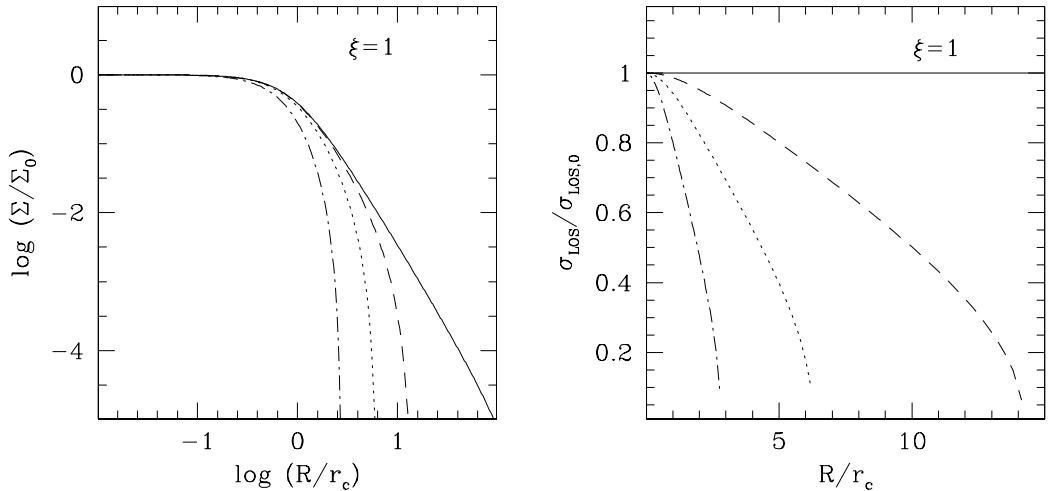}
 \caption{Projected surface-density and LOS velocity-dispersion
     profiles of $\xi=1$ MOND models with $W_0=3$ (dot-dashed lines),
   $W_0=6$ (dotted lines) and $W_0=9$ (dashed lines). The $\xi=1$ MOND
   isothermal sphere profiles (Milgrom 1984) are shown for comparison
   in both panels as solid lines.}
\label{w}
\end{figure*}

We describe here the structural and kinematic properties of the models
presented in the previous Section.  For a given interpolating function
$\mu$, a given choice of $W_{0}$ and $\xi$ corresponds to a different
model. Here (and below, when not specified otherwise) we adopt the
interpolating function (Famaey \& Binney~2005)
\begin{equation}
\mu(y)={y\over 1+y}.
\label{eq_mu}
\end{equation}
The knowledge of the cluster mass $M$ allows us to determine $\rc$,
$\rho_{0}$ and $\sigmaK$ using the relations
\begin{equation}
\rc=\left[\frac{GM}{9 a_{0} \xi I(W_{0},\xi)}\right]^{\frac{1}{2}},
\end{equation}
\begin{equation}
 \rho_{0}=\frac{9 \xi a_{0}}{4\pi G \rc}, 
\end{equation}
and
\begin{equation}
\sigmaK=\left[\frac{GM\xi a_{0}}{9 I(W_{0},\xi)}\right]^{\frac{1}{4}},
\end{equation}
where
\begin{equation}
I(W_{0},\xi)\equiv\int_{0}^{\rtilde_{\rm
    t}}\rtilde^{2}\tilde{\rho}~d\rtilde,
\end{equation}
with $\rtilde_{\rm t}\equiv\rt/\rc$.  The above expressions can be
used to transform the dimensionless quantities to physical ones.

In Fig.~\ref{pot} we plot the density profile (top panel), the
gravitational potential (central panel) and the acceleration modulus
(bottom panel) of a Newtonian King model (solid curves) with
$W_{0}=6,~\log(M/M_{\odot})=5$ and $\rc=5~{\rm pc}$ and, for
comparison, the same quantities for a MOND model (dashed curves)
having the same mass and reproducing the same surface-density profile
($W_{0}=9.5,~\xi=0.53$). As can be seen, the MOND model has a
potential well $\sim$3 times deeper than the Newtonian model's one.
Correspondingly, the MOND model has a substantially higher
acceleration than the Newtonian model at all radii (see the bottom
panel of Fig.~\ref{pot}). It must be noted that in the considered
example, which is representative of a low surface-density cluster, the
Newtonian acceleration is at all radii significantly lower than
$\azero$, so the effects of MOND are large at any distance from
 the cluster centre.

In Figs.~\ref{xi} and \ref{w} the surface-density and LOS
velocity-dispersion profiles for a set of models with different
choices of $W_0$ and $\xi$ are shown. From Fig.~\ref{xi} it is
apparent that MOND models predict steeper density and
velocity-dispersion profiles with respect to Newtonian models having
the same values of $W_{0}$.  Indeed, the larger acceleration predicted
by MOND at large distances from the cluster centre implies a steeper
increase of the effective potential $\psi$ that reaches zero at a
smaller radius ($r_{\rm t}$) with respect to the Newtonian case. MOND
models tend to Newtonian ones when large values of $\xi$ are
considered (see Fig. \ref{xi}): the Newtonian case can be reproduced
for $\azero\to 0$ (i.e. no characteristic acceleration). In this
limit, $\xi\rightarrow \infty$, $\mu\to 1$, and
equation~(\ref{poiss_eq}) approaches Poisson's equation. $W_0$
measures the depth of the gravitational potential well, so --- like in
Newtonian King models --- as $W_{0}\rightarrow \infty$, the escape
velocity goes to infinity at any distance from the cluster centre, the
velocity distribution tends to the Maxwellian distribution, and the
models approach the isothermal sphere (see Fig.  \ref{w}).

We note that, with the exception of the isothermal sphere, in our MOND
models $\sigmaLOS$ goes to zero when $r$ approaches the tidal
  radius (see right panels of Fig.~\ref{xi} and Fig.~\ref{w}), as it
must for the system to be bounded.  In fact, as $r\rightarrow
r_{\rm t}$ the effective potential $\psi \rightarrow 0$ and the
velocity distribution defined in equation~(\ref{df_eq}) tends to a null
width function. This is not the case for the isothermal sphere whose
distribution function has the same shape regardless of the potential,
and it is not truncated, though having finite mass in MOND (Milgrom~1984).

\section{Application to globular clusters in the outer Galactic halo}
 
\begin{figure*}
 \includegraphics[width=15.cm]{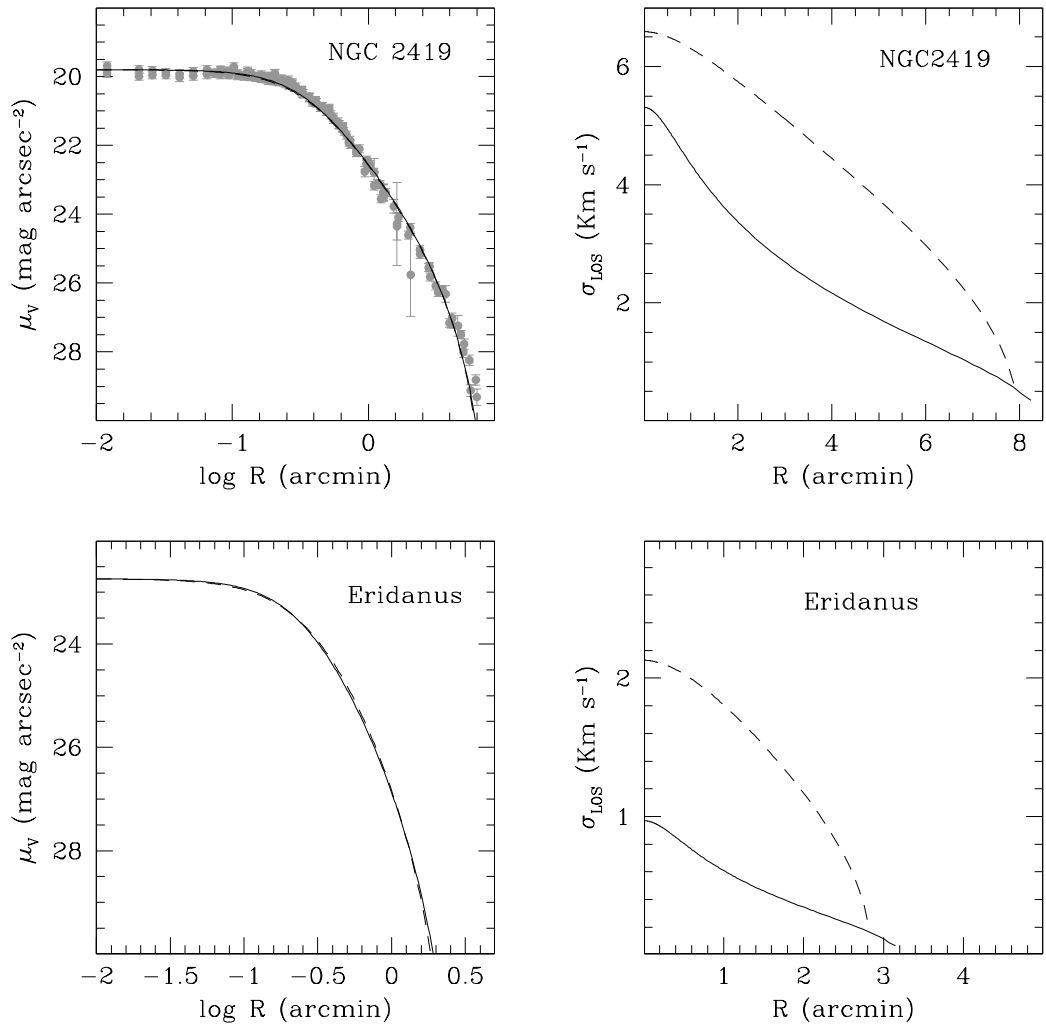}
 \caption{Surface-brightness (left panel) and LOS velocity-dispersion
   (right panel) profiles according to MOND (dashed lines) and
   Newtonian (solid lines) models for the globular clusters NGC2419
   and Eridanus.  The surface-brightness measures for NGC2419 from
   Trager et al. (1995) are overplotted in the upper left panel with
   grey points.}
\label{gc1}
\end{figure*}

 \begin{figure*}
 \includegraphics[width=15.cm]{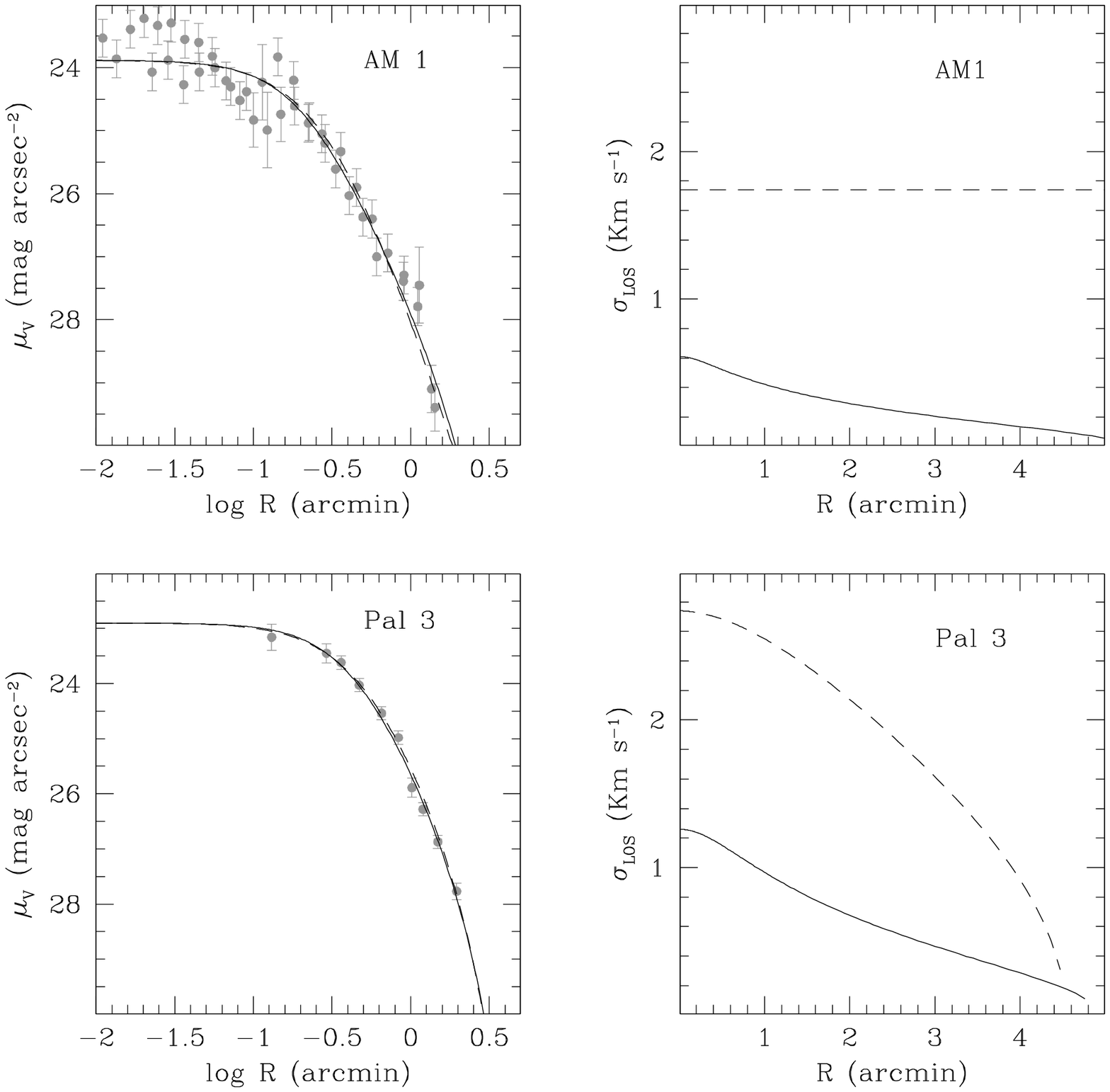}
 \caption{Same as Fig. \ref{gc1}, but for the globular clusters AM 1
   and Pal 3.}
\label{gc2}
\end{figure*}

\begin{figure*}
 \includegraphics[width=15.cm]{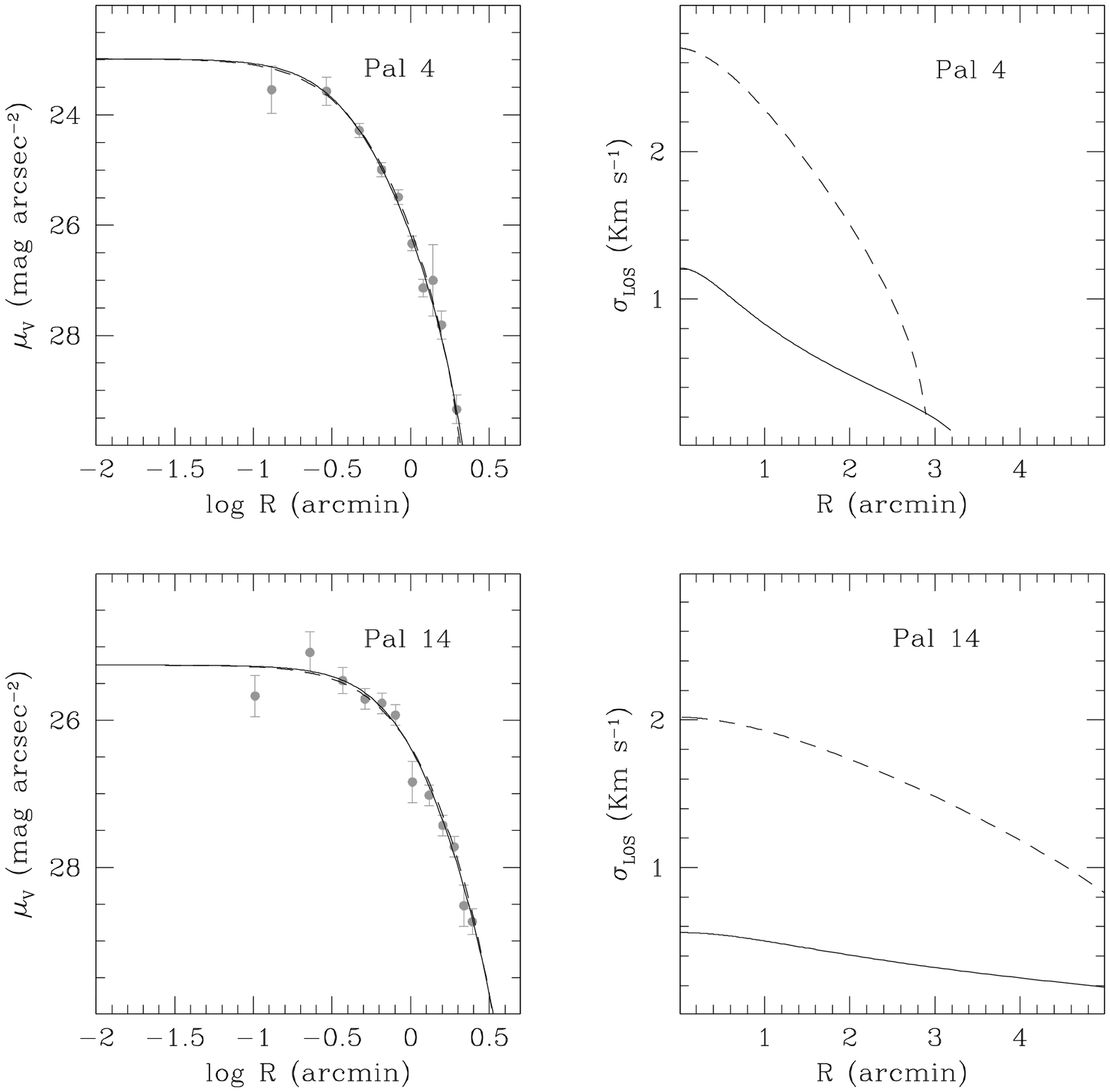}
 \caption{Same as Fig. \ref{gc1}, but for the globular clusters Pal 4
   and Pal 14.}
\label{gc3}
\end{figure*}

The examples shown in the previous Section suggest that the shape of
the velocity-dispersion profile of a GC of given density profile can
be quite different in MOND and Newtonian gravity. This makes the
velocity-dispersion profiles of GCs a useful tool for
distinguishing between MOND and Newtonian gravity.
Here we apply our models to specific GCs, which may be used to test
the MOND theory by combining measures of surface-brightness and LOS
velocity-dispersion profiles.  The models presented in this paper are
valid in the external acceleration range $0<\gext\ll a_{0}$ (see
Section~\ref{ext_sec}). In spite of this limitation, a number of GCs
belonging to the Galactic halo satisfy this constraint. Unfortunately,
although the surface-brightness profiles of these clusters are well
measured (McLaughlin \& van der Marel 2005, hereafter MvdM05), to date 
none of them have accurate velocity-dispersion profiles. 
A first attempt at estimating the velocity-dispersion
profile of NGC2419 has been done recently by Baumgardt et al. (2009).

Here we compare the velocity-dispersion profiles predicted by the
Newtonian and MOND models that reproduce the surface-brightness
profiles of six GCs located in the outer halo of the Milky Way (at
distances $>50$ kpc from the Galactic centre).  These clusters (listed
in Table~1) are subject to an external Galactic acceleration
$\gext \lsim 3\times 10^{-11} \ms \ll a_{0}$, so they are fully in the
regime of validity of our models.

\subsection{MOND and Newtonian velocity-dispersion profiles for given mass and size}

Though the masses and physical sizes of GCs are known only with
non-negligible uncertainties, it is useful to discuss first the
idealized case in which these quantities are given.  In other words,
for each object we fix a value of the mass-to-light ratio $M/L$ and of
the cluster distance, valid for both Newtonian and MOND models, which
is used to convert surface brightness in surface mass density.  We
defer to Section~\ref{sec:moverl} a discussion of the effects of
varying $M/L$ and distance. For Newtonian models we adopt the central
dimensionless potentials $W_0$, core radii $\rc$ and masses $M$ from
MvdM05.  For MOND models, we adopt the cluster mass estimated by
MvdM05 and search for the values of $W_{0}$ and $\xi$ that best
reproduce the surface-density profile of the corresponding
  Newtonian models.

In Table~1 the main structural properties derived for the six
considered clusters are summarized.  For Eridanus (which is not
included in the catalog of MvdM05) we adopted the concentration, core
radius, visual absolute magnitude given by Harris (1996) and
estimated its mass by adopting a mass--V-band-luminosity ratio
$M/\LV=1.892$ (the same value adopted by MvdM05 for Pal 4, which
has similar age and metallicity; Catelan 2000).  In both Newtonian
and MOND cases, the central LOS velocity dispersion $\sigma_{\rm
  LOS,0}$ has been calculated by using equation~(\ref{los_eq}).

Figures \ref{gc1}, \ref{gc2} and \ref{gc3} show the surface-brightness
profiles (left panels) and velocity-dispersion profiles (right panels)
of the Newtonian and MOND models of the six considered GCs. In the
left panels we report also, where available, surface-brightness
measures from Trager, King \& Djorgovski (1995). As can be noted, MOND
predictions significantly differ from Newtonian ones: in particular,
it is apparent that the {\it shape} of MOND and Newtonian LOS
velocity-dispersion profiles are remarkably different. As expected,
MOND models predict a larger velocity dispersion along the entire
cluster extent with respect to Newtonian models with the same mass.
The physical reason at the basis of this result is that, as showed in
Fig.~\ref{pot}, the same observed surface mass density profile is
fitted by MOND models with a deeper potential well and larger core
radii.  As a consequence, at a given radius, the corresponding escape
velocity is larger and according to equation~(\ref{df_eq}) the width
of the velocity distribution turns out to be larger.  The same
qualitative results have been obtained by Moffat \& Toth~(2008) and
Haghi et al.~(2009), who adopted different methods to calculate the
velocity-dispersion profiles of the sample of GCs indicated by BGK05.
 
We note that the MOND $\sigmaLOS$ profiles of our clusters are in
general decreasing functions of radius and $\sigmaLOS$ goes to zero
when $r$ approaches the truncation radius, consistent with the fact
that the no stars can cross the system's boundary (see
Sect. \ref{res_sect}). An exception is the GC AM1, for which the
predicted MOND velocity-dispersion profile in AM1 is flat along the
entire cluster extent. The surface-brightness profile of this cluster
is indeed well fitted by the MOND isothermal sphere model
(Milgrom~1984) corresponding to its mass.  The maximum difference in
the velocity dispersion predicted by MOND and Newtonian theories
ranges from $1.2\kms$ (Eridanus) to $2.4\kms$ (NGC2419), well above
the accuracy currently achievable with high-resolution spectroscopic
analyses.

\begin{table*}
 \centering
 \begin{minipage}{140mm}
   \caption{Structural parameters of the six considered outer halo
     globular clusters.}
   \begin{tabular}{@{}lccccccccccr@{}}
  \hline
        &                     &                 & & \multicolumn{3}{c}{Newtonian} & \multicolumn{4}{c}{MOND} &\\
   Name & $\log (M/M_{\odot})$ & $\gext/\azero$ & $M/\LV$ & $W_{0}$ & $\rc$ & $\sigma_{\rm LOS,0}$ & $W_{0}$ & $\xi$ & $\kappa$ & $\sigma_{\rm LOS,0}$ & $\Delta\sigma_{\rm LOS}^{max}$\\
        &                     &                 & $M_{\odot}/L_{V,\odot}$  &      &  (pc) &   $({\rm km}~{\rm s}^{-1})$      &         &     &  &  $({\rm km}~{\rm s}^{-1})$       & $({\rm km}~{\rm s}^{-1})$\\
 \hline
 NGC2419  & 5.95 & 0.12 & 1.903 & 6.5 & 8.41 & 5.31 & 10.0 & 1.25  & 11.470 & 6.59 & 2.43\\
 Eridanus & 4.21 & 0.11 & 1.892 & 5.3 & 6.56 & 0.97 &  8.5 & 0.10  &  0.118 & 2.13 & 1.23\\
 AM1      & 4.01 & 0.09 & 1.868 & 6.6 & 7.16 & 0.61 & $\infty$ & 0.04 &  0.028 & 1.74 & 1.69\\
 Pal 3    & 4.65 & 0.11 & 1.869 & 5.3 & 10.68 & 1.26 & 8.5 & 0.10  &  0.118 & 2.74 & 1.58\\
 Pal 4    & 4.58 & 0.10 & 1.892 & 4.5 & 12.16 & 1.21 & 7 & 0.09    &  0.079 & 2.70 & 1.51\\
 Pal 14   & 4.09 & 0.16 & 1.885 & 4.3 & 19.60 & 0.56 & 6.5 & 0.02  &  0.004 & 2.02 & 1.46\\
\hline
\end{tabular}
\end{minipage}
\end{table*}

\subsection{Dependence on mass-to-light ratio and  distance}
\label{sec:moverl}

\begin{figure}
 \includegraphics[width=8.7cm]{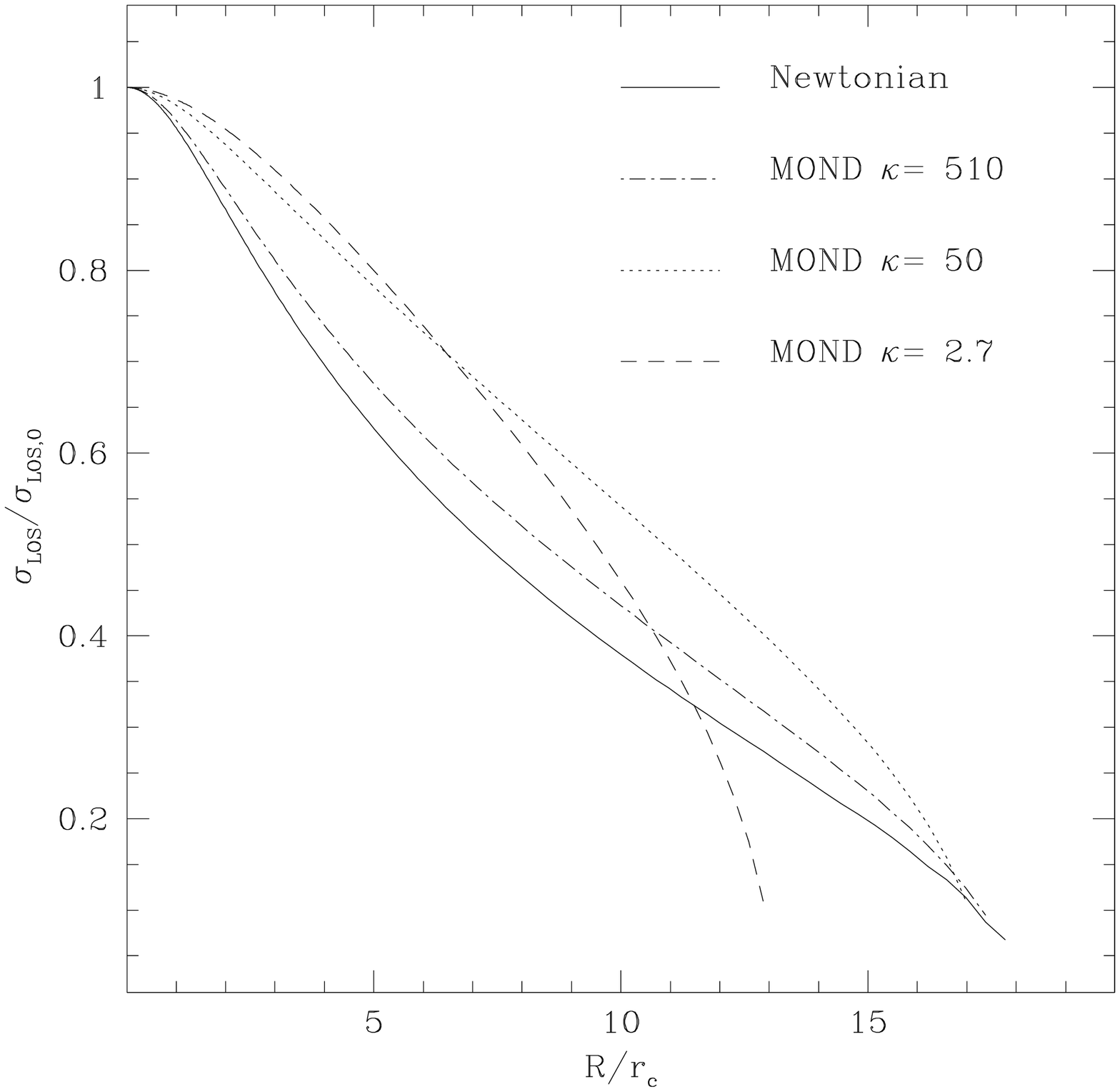}
  \caption{LOS velocity-dispersion profiles of a Newtonian King model
    with $W_{0}=6$ (solid line) compared with a family of MOND models
    with different values of $\kappa$, but with the same shape of the
    surface-brightness profile.}
\label{par2}
\end{figure}

While the {\it shape} of the density and velocity-dispersion profiles
of Newtonian King models does not depend on the structural parameters
(mass and core radius), this is not the case for the corresponding
MOND models. In these cases, the shape of the profiles varies for
varying $\rc$ and/or $M$. In other words, while Newtonian King models
of given $W_0$ can be rescaled to represent systems of arbitrary
values (in physical units) of $\rc$ and $M$, a MOND model of given
$W_0$ and $\xi$ represents only systems such that
\begin{equation}
\kappa\equiv\frac{GM}{a_{0}\rc^2}=9\xi I(W_{0},\xi)
\label{eq:kappa}
\end{equation}
(see also Nipoti, Londrillo \& Ciotti 2007c, for a detailed discussion
of scaling of MOND models). For each value of $\kappa$ there is a
unique pair of parameters ($W_0,~\xi$) that reproduces a given shape
of the surface-brightness profile. The shape of the corresponding
velocity-dispersion profiles is different for different values of
$\kappa$.  This is illustrated in Fig. \ref{par2}, where the
velocity-dispersion profile of a Newtonian model with $W_{0}=6$ is
compared with a family of MOND models that share the same
surface-brightness profile, but have different values of $\kappa$.
Here the LOS velocity dispersion and the radius are normalized to
$\sigma_{\rm LOS,0}$ and $\rc$ , respectively, to highlight the
different shape of the profiles.  It is evident that while small
values of $\kappa$ produce steep profiles, the MOND profiles approach
the Newtonian one for increasing $\kappa$. In fact, the larger
$\kappa$ the larger the internal acceleration of the cluster, 
  which eventually exceeds $\azero$ over most of the cluster extent.

In a practical application, once the surface-brightness profile of a
given cluster is known, a unique pair of parameters ($W_0,~\xi$)
that reproduces the shape of the observed surface-brightness profile
can be determined only when an estimate of $M$ and $\rc$ is provided.
On the other hand, the uncertainties on the mass-luminosity ratio
and distance produce an uncertainty on $M$ and $\rc$ and,
consequently, on the predicted shape of the velocity-dispersion
profile.

\begin{figure}
 \includegraphics[width=8.7cm]{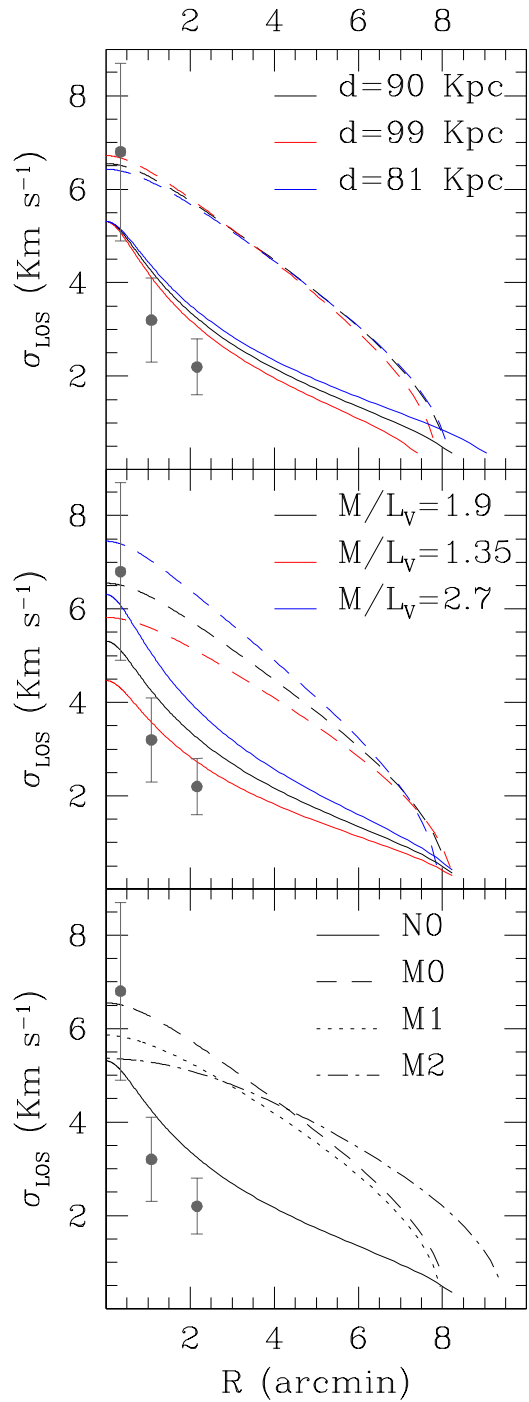}
  \caption{LOS velocity-dispersion profiles for NGC2419. In the top
    panel Newtonian models (solid lines) are compared with MOND models
    (dashed lines) with different assumptions on the cluster distance
    $d$.  In the central panel Newtonian models (solid lines) are
    compared with MOND models (dashed lines) with different
    assumptions on the mass-to-light ratio.  In the bottom panel the
    Newtonian model N0 (solid line) is compared with MOND models M0
    (dashed line), M1 (dotted line) and M2 (dot-dashed line) differing
    in the choice of the interpolating function $\mu$ (see Table~2).
    The data by Baumgardt et al.~(2009) are overplotted in the three panels. }
\label{par}
\end{figure}

To illustrate this issue, in Fig.~\ref{par} we show the
velocity-dispersion profiles of the models that reproduce the
surface-brightness profile of NGC2419 assuming a different mass
(central panel) and a different core radius (top panel). In
particular, we let the cluster mass vary by
$\Delta~\log\,(M/M_{\odot})=\pm 0.15$, with respect to the reference
value $\log\,(M/M_{\odot})=5.95$ (therefore exploring the cases
$M/\LV=2.7$ and $M/\LV=1.35$, beside the reference case $M/\LV=1.9$),
and we let the cluster core radius vary by $\Delta~\rc=\pm0.1~\rc$
(exploring the cases of cluster distance $d=81$ kpc and $d=99$ kpc,
beside the reference case $d=90$ kpc). The corresponding overall
cluster velocity dispersions (calculated by integrating $\sigmaLOS$
over the entire cluster extent) are listed in Table~2.  While a change
in the core radius does not significantly affect either the shape or
the central value of the velocity-dispersion profile, an even
relatively small variation of $M$ significantly alters the overall
value of the velocity dispersion.

It must be noted that even comparing the highest $M/L$ (and smallest
distance) Newtonian model with the lowest $M/L$ (and larger distance)
MOND model, the resulting LOS velocity-dispersion profiles are
remarkably different and thus observationally distinguishable.

\subsection{Dependence  on the interpolating function $\mu$}

An additional uncertainty is related to the adopted form of the
interpolating function $\mu$ (appearing in equation~\ref{eqMOND}),
which is not constrained theoretically, except for the asymptotic
behaviour (see Section~\ref{intro}). The choice of the functional form
of $\mu$ determines the behaviour of the MOND acceleration strength of
the MOND effects in the intermediate acceleration regime (when $g$ is
of the order of $\azero$), thus changing the overall shape of the
velocity-dispersion profile.

In the bottom panel of Fig.~\ref{par} we show the velocity-dispersion
profiles of the models that reproduce the surface-brightness profile
of NGC2419 assuming different forms of the interpolating function
$\mu$ (see Table~2), but keeping fixed $M$ and $\rc$. As expected,
both the morphology of the velocity-dispersion profile and its average
value depend on $\mu$.  However, the effect of varying $\mu$ is
typically small if we limit ourselves to standard proposal such as
equation~(\ref{eq_mu}) (model M0) or Milgrom's (1983)
$\mu(y)=y/\sqrt{1+y^2}$ (model M1), and we exclude unrealistic cases
such as the step function adopted in model M2.

\subsection{Comparison with velocity-dispersion measures}

\begin{table}
 \centering
  \caption{Predicted overall velocity dispersion for different MOND and Newtonian models of NGC2419.}
   \begin{tabular}{@{}lcccr@{}}
  \hline
   Model & $\mu(y)$ 	     & $\log\,(M/M_{\odot})$ & $\rc$ & $<\sigmaLOS>$ \\
         &          	     &  		   &  (pc)   &   $(\kms)$ \\
 \hline
 N0      &      1      	     & 5.95 & 8.41  & 4.40\\
 Nm-     &      1     	     & 5.80 & 8.41  & 3.70\\
 Nm+     &      1      	     & 6.10 & 8.41  & 5.23\\
 Nr-     &      1     	     & 5.95 & 7.57  & 4.64\\
 Nr+     &      1    	     & 5.95 & 9.25  & 4.20\\
 M0      & $y/(1+y)$  	     & 5.95 & 9.50  & 6.18\\
 M1      & $y/\sqrt{1+y^{2}}$ & 5.95 & 8.50  & 5.56\\
 M2      & $\max(y,1)$        & 5.95 & 7.96  & 5.22\\
 Mm-     & $y/(1+y)$  	     & 5.80 & 10.28 & 5.51\\
 Mm+     & $y/(1+y)$  	     & 6.10 & 9.23  & 6.93\\
 Mr-     & $y/(1+y)$  	     & 5.95 & 8.54  & 6.28\\
 Mr+     & $y/(1+y)$  	     & 5.95 & 11.08 & 6.07\\
\hline 
\end{tabular}
\end{table}

We have seen that, at least for the representative case of NGC2419,
while the same overall velocity dispersion can be predicted by both
Newtonian and MOND models with different choices of the mass-to-light
ratio or of the interpolating function (compare, e.g., model Nm+ with
models M1, M2, and Mm- in Table~2), MOND profiles can be always easily
distinguished from Newtonian ones.  The analysis of the shape of the
velocity-dispersion profile represents therefore a robust method to
discriminate between the two gravity theories.  This method is indeed
less sensitive to the errors on the cluster mass with respect to the
simple comparison between the overall cluster velocity dispersion
proposed by BGK05.  Moreover, the approach suggested by these authors
favours low-mass GCs whose internal acceleration is lower than
$\azero$ at any distance from the cluster centre. The low mass of
these clusters, together with their large distances, imply a poor
efficiency in measuring radial velocities for a meaningful sample of
stars.

Of course, measuring the velocity-dispersion profile is
observationally more challenging than estimating the overall velocity
dispersion, but these kinds of measures are becoming feasible even for
relatively distant GCs. In Fig.~\ref{par} we overplot the three
velocity-dispersion measures obtained by Baumgardt et al.~(2009)
from spectroscopic observations of 40 stars of NGC2419. Though
  there are large uncertainties due to the poor statistics, the
overall trend defined by these observations appears hard to reconcile
with MOND, at least under the considered assumption of spherical
symmetry and isotropic velocity distribution. This preliminary result
confirms that NGC~2419 might be a crucial object to test MOND (as also
suggested by Baumgardt et al.~2009) and strongly encourages future
studies of this object combining higher resolution observations of a
larger number of stars of NGC2419 as well as a systematic study of the
possible effects of orbital anisotropy, rotation and deviation from
spherical symmetry. Among these effects, that of orbital
  anisotropy is likely the most important, because the shape of the
  LOS velocity-dispersion profile can depend significantly on the
  distribution of stellar orbits. In particular, a radially
  anisotropic system is expected to have centrally steeper $\sigmaLOS$
  profile than an isotropic system with the same spatial
  distribution. Therefore, it is worth investigating whether radially
  anisotropic MOND models can be reconciled with the velocity
  dispersion data for NGC2419. We address this question in the
  next Section.

\subsection{Radially anisotropic models of NGC2419}
\label{secanis}

Ideally, to explore the effect of orbital anisotropy on the
  $\sigmaLOS$ profiles of GCs, one would need self-consistent
  anisotropic MOND models derived from the distribution
  function. Constructing such models is beyond the purpose of the
  present work: here we perform a preliminary analysis based on
  the numerical integration of the Jeans equations.  We follow the
  standard procedure (Binney \& Mamon~1982), but with the MOND
  gravitational field replacing the Newtonian field.  In practice, we
  take the spherically symmetric density distribution of the King
  model of NGC2419 (parameters in Table~1), rescale it for the
  assumed value of $M/\LV$, and compute the MOND field generated by
  this density distribution using equation~(\ref{eqMONDsph}). We then
  solve the Jeans equations assuming an anisotropy-parameter profile
  $\beta(r)\equiv 1-\sigma^2_{\rm t}(r)/2\sigma^2_{r}(r)$, where
  $\sigma^2_{\rm t}$ and $\sigma^2_{r}$ are the tangential and radial
  components of the velocity dispersion tensor. Finally, we obtain the
  $\sigmaLOS(r)$ by deprojecting $\sigma^2_{r}(r)$. For comparison, we
  also obtain $\sigmaLOS$ profiles of anisotropic Newtonian models
  using the same procedure.

\begin{figure}
 \includegraphics[width=8.7cm]{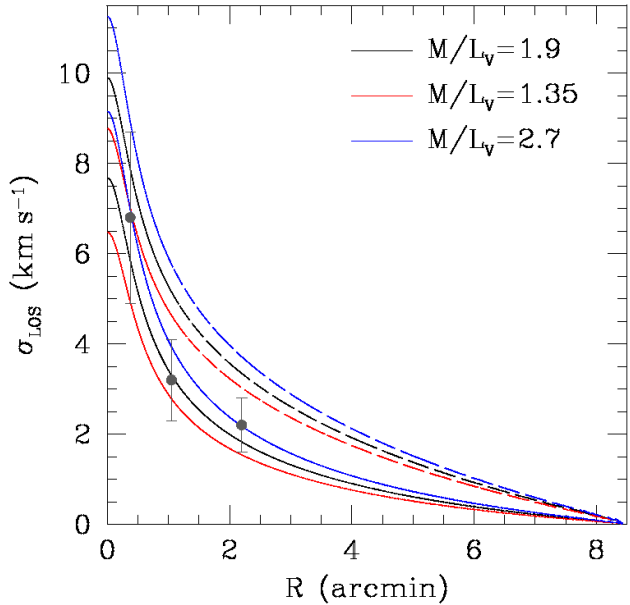}
 \caption{Same as the middle panel of Fig.~\ref{par}, but for radially
   anisotropic (Osipkov-Merritt) Newtonian (solid lines) and MOND
   (dashed lines) models of NGC2419.}
\label{anis}
\end{figure}

  As stressed in the Introduction, the Jeans-equations approach does
  not guarantee that the obtained models are self-consistent. However,
  we can at least use some necessary conditions for consistency (e.g.,
  Ciotti \& Pellegrini 1992; An \& Evans 2006; Ciotti \&
  Morganti~2009) to exclude unphysical $\beta(r)$. We note that these
  necessary conditions, though derived in the context of Newtonian
  gravity, apply to our self-gravitating MOND models, because each of
  this models can be formally interpreted as a non-self-gravitating
  distribution of tracer stars in the presence of a dominant mass
  distribution having the same gravitational potential as the MOND
  potential of the cluster. An \& Evans (2006) show that a necessary
  condition for consistency\footnote{This condition applies to a
    distribution of tracer stars in a potential well with finite depth
    (Evans, An \& Walker 2009), as is the case for the MOND potentials
    of our models (see middle panel in Fig.~\ref{pot}).} is
  $\beta(0)\leq \gamma/2$, where $\gamma\equiv-\lim_{r\to 0}d \ln \rho
  /\d \ln r$ is the central logarithmic slope of the stellar density
  distribution. For NGC2419 $\gamma\sim 0$, so models with
  $\beta(0)\gsim 0$ are inconsistent, implying that spherical,
  radially anisotropic models with $\beta$ independent of radius are
  unphysical. We then consider Osipkov-Merritt (hereafter OM) models
  (Ospikov 1979; Merritt 1985), which are isotropic in the centre and
  radially anisotropic at large radii, having
  \begin{equation}
    \beta(r)= {r^2\over r^2+\ra^2}, 
\end{equation}
where $\ra$ is the anisotropy radius.  A necessary condition
for the consistency of OM models is that $(r^2+\ra^2)\rho(r)$ is a
non-increasing function of radius (Ciotti \& Pellegrini~1992). In the
considered model of NGC2419 this condition is satisfied for $\ra\gsim
9.43\,{\rm pc} \simeq 1.12\rc$, where $\rc$ is the core radius given
in Table~1. In Fig.~\ref{anis} we plot the $\sigmaLOS$ profile for
maximally radially anisotropic ($\ra=1.12\rc$) MOND and Newtonian OM
models of NGC2419 for the three reference values of $M/\LV$ adopted in Fig.
\ref{par}: as
expected, radially anisotropic models predict higher $\sigmaLOS$ in
the centre and lower $\sigmaLOS$ in the outer regions than the
corresponding isotropic models (middle panel in Fig.~\ref{par}). While
radially anisotropic Newtonian models reproduce well the observational
data from Baumgardt et al.~(2009), MOND anisotropic models still tend
to predict too high velocity dispersion.  

Given the large uncertainties in the observational data, it
is not excluded that the velocity-dispersion profile of NGC2419 can be
reproduced by a MOND radially anisotropic model with low stellar
mass-to-light ratio. However, it must be stressed that the models
plotted in Fig.~\ref{anis} are extreme cases: they satisfy only the
{\it necessary} condition for consistency, so they may have to
be excluded as unphysical or unstable.

\section{Summary and conclusions}

We have presented self-consistent dynamical models of stellar systems in
MOND that can be used to study GCs in the outer Galactic regions.
These models, which are the analogues of the Newtonian King models,
are able to reproduce the observed surface-brightness profiles of GCs
and can be used to predict their projected velocity-dispersion
profiles in MOND. By comparing our models with the corresponding
Newtonian ones, we found that it is impossible to reproduce
simultaneously the same density and velocity-dispersion profiles with
both gravitational theories regardless of any choice of the free
parameters of the models. This indicates GCs are one of the best
laboratories to test the gravity theory in the low-acceleration
regime, as already suggested by various authors (Scarpa, Marconi
  \& Gilmozzi 2003; BGK05; Scarpa et al. 2007; Moffat \& Toth 2008;
  Haghi et al. 2009; Lane et al. 2009).

Of course, the same effect produced by MOND on the velocity-dispersion
profile can by reproduced by an $ad~hoc$ distribution of dark
matter. The presence of dark-matter halos in GCs is predicted by some
theories of GC formation and evolution (see Mashchenko \& Sills 2005
and references therein) and its observational evidence is still matter
of debate (Moore 1996; Forbes et al.  2008). Thus, the detection of
velocity-dispersion profiles in agreement with the predictions of MOND
would not necessarily be a problem for the dark-matter paradigm.  On
the other hand, it must be stressed that the detection of
velocity-dispersion profiles expected on the basis of Newtonian
dynamics (without dark matter) could invalidate MOND.

Testing MOND by using nearby GCs has been already attempted by Scarpa
et al. (2003) who found that their velocity-dispersion profiles
deviate from the prediction of Newtonian dynamics (without dark
matter) at large distance from their centres. However, the nearby
clusters analysed by these authors experience an external acceleration
due to the Milky Way gravitational field that is larger than the
critical acceleration $a_{0}$ (BGK05; Moffat \& Toth 2008).  BGK05
indicated a sample of eight GCs where the predictions of MOND and
Newtonian theories on the overall LOS velocity dispersion
significantly differ. However, the method proposed by these authors is
very sensitive to the adopted mass and distance of these clusters.
Given the large distances and low mass of these clusters, it is very
difficult to observe the significant number of cluster member stars
necessary to an accurate estimate of the velocity dispersion. In this
regard, the difference between the MOND and Newtonian predictions
estimated by these authors never exceeds $\Delta \sigmaLOS<1.3\kms$.
Given the best accuracy achievable by the current observing facilities
($\sim 0.5\kms$), it is hard to reach a firm conclusion on the
validity of MOND using this approach. Nevertheless, the first results
of the application of this method to Pal 14 suggests that the observed
kinematic of this system might be a problem for MOND (Jordi et
al. 2009).  

A more robust test is to compare the observed shape of the
velocity-dispersion profile with the prediction of Newtonian and MOND
models (see also Moffat \& Toth 2008; Haghi et al. 2009). We
demonstrated that this method always allows an unambiguous distinction
between Newtonian and MOND scenarios even when large uncertainties on
the cluster mass and core radii are present. The best target GC, for
which this approach is expected to be particularly efficient, is
NGC2419. Indeed, although not included in the list of BGK05, this
cluster is the one with the largest absolute difference in the
predicted velocity-dispersion profile by Newtonian and MOND
models. Moreover, it is not significantly affected by Galactic tidal
effects (Gnedin \& Ostriker 1999) that can alter the shape of the
velocity-dispersion profile in the outermost regions (Johnston,
Sigurdsson \& Hernquist 1999) and is massive enough to ensure a large
number of stars at magnitudes easily reachable by the current
generation of spectrographs. Recently Baumgardt et al. (2009) measured
the velocity dispersion at three different radii in this cluster.  The
two outermost data points are significantly lower than the
velocity-dispersion profiles predicted by our isotropic MOND
  models of NGC2419, at least for stellar mass-to-light ratios in the
  range $1.35 \lsim M/\LV \lsim 2.7$. Only assuming quite strong
  radial orbital anisotropy and lower stellar mass-to-light ratio
  the velocity dispersion predicted by MOND can be reconciled with the
  observed data of NGC2419. Reproducing the observed kinematics of
NGC2419 represents a challenge for MOND, although better data sets
(larger number of stars and higher spectral resolution) and more
sophisticated modelling are needed.

A limitation of the present study is that our models are spherically
symmetric, non rotating and with isotropic velocity distribution.  The
assumption of spherical symmetry and absence of significant rotation
is in general justified by the round appearance of GCs. 
In general, non-sphericity can be a problem in the determination 
of surface brightness and velocity
dispersion profiles when large ellipticities are present (Perina et al. 2009). 
This effect should have only a minor impact in GCs which have generally 
symmetric density contours, at least within few core radii.
The amount of
anisotropy in many GCs is estimated to be relatively small (Ashurov \&
Nuritdinov 2001), though a non-negligible degree of anisotropy is
likely to be present in few GCs (Meylan \& Heggie 1997 and references
therein). Orbital anisotropy is also predicted by N-body simulations
as a result of both primordial and evolutionary reasons (Giersz 2006).
These effects are expected to be at least partially erased in Galactic
GCs by the strong tidal interaction with the Milky Way which removes
the initial velocity anisotropies and angular momentum making them
more spherical (Goodwin 1997). A preliminary exploration of the effects of
an extreme radial anisotropy shows a significant
degeneracy between gravity law and orbital anisotropy. Nevertheless, the MOND
and Newtonian models can be distinguished when high precision data are
available. 
Another possible complication can be due to the presence of a significant
fraction of unresolved binaries which can inflate the observed velocity 
dispersions (Cote et al. 2002). For instance, in NGC2419 the orbital velocity of a pair of
equal-mass 0.8 $M_{\odot}$ stars separated by $a\leq 14~AU$ could be as high
as 10 $Km~s^{-1}$, remaining stable against collisional disruption (Hills 1984). 
A significant fraction of binaries can therefore increase the
observed velocity dispersion by few $Km~s^{-1}$. This effect is particularly
important in the central part of the cluster where binaries preferentially
sink as a result of mass segregation. 
This could explain why the measured central velocity dispersion in
NGC2419 seems higher than the prediction of both Newtonian and MOND models.
Note however that, in the case of NGC2419, the "binary-corrected" velocity dispersion would 
stray even more from the prediction of MOND models in the external region of the 
cluster.
Given these uncertainties, a
systematic exploration of the effects on the cluster kinematics of
orbital anisotropy, rotation and deviation from spherical symmetry in
general would be valuable also in MOND as well as in Newtonian
dynamics (see Bertin \& Varri~2008, and references therein). 

\section*{Acknowledgments}

This research has been supported by the Instituto de
Astrofisica de Canarias.
We warmly thank Michele Bellazzini and Luca Ciotti for helpful
discussions. We also thank the anonymous referee for his/her helpful comments
and suggestions.

\appendix

\section{Numerical integration}

We describe briefly the procedure used to numerically integrate
equation~(\ref{poiss_eq}), from $\rtilde=0$ to a given radius
$\rtilde$, to obtain $W(\rtilde)$. Equation~(\ref{poiss_eq}) can be
written as
\begin{equation}
\left[\frac{2}{\rtilde}\frac{\d W}{\d\rtilde}+\frac{\d^{2}W}{\d\rtilde^{2}}\right]
\mu\left(\xi\left|{\d W \over \d \rtilde}\right|\right)
-\frac{\d W}{\d\rtilde}\xi\frac{\d^{2}W}{\d\rtilde^{2}}\mu^{\prime}\left(\xi\left|{\d W \over \d \rtilde}\right|\right)
 =  -9\tilde{\rho}
\end{equation}
where
\begin{equation}
\mu^{\prime}(y)\equiv\frac{d\mu(y)}{dy}.
\end{equation}
The term $\d^{2}W/\d\rtilde^{2}$ diverges at the centre, but, for any
choice of the MOND interpolating function $\mu$, substitutions of
power series show that
\begin{equation}
\lim_{\rtilde\rightarrow0}\frac{\d^{2}W}{\d\rtilde^{2}}=-\sqrt{\frac{3}{4\xi}}\rtilde^{-\frac{1}{2}}
\end{equation}
whose integral converges and admits the exact solution
\begin{eqnarray*}
&&\left. \frac{\d W}{\d\rtilde} \right
\vert_{\rtilde\rightarrow0}=-\sqrt{\frac{3}{\xi}}\rtilde^{\frac{1}{2}},\\
&&W \vert_{\rtilde\rightarrow0}=W_{0}-\sqrt{\frac{4}{3\xi}}\rtilde^{\frac{3}{2}}.
\end{eqnarray*}
We used the above equations to calculate the dimensionless potential
and its derivatives at the centre.  For the subsequent integration
steps we adopted a radial step $\Delta \rtilde$ such that
\begin{equation}
 \max \left( \frac{\d W}{\d \rtilde} \Delta \rtilde,\frac{\d^{2}W}{\d\rtilde^{2}}\Delta \rtilde\right)<0.01,
\end{equation}
in order to ensure a negligible numerical error in the integrations.

\bsp

\label{lastpage}

\end{document}